\documentclass[prl,twocolumn,superscriptaddress]{revtex4-1}
\usepackage{bm}
\usepackage{amsmath,amssymb,amsfonts}%
\usepackage{siunitx}%for SI unit input
\usepackage{chemarrow}% for bi-directional arrow with text over and below
\usepackage{nicefrac}
\usepackage{braket}
\usepackage{graphicx}
\usepackage{xcolor}

\usepackage{hyperref}
\hypersetup{
     colorlinks   = true,
     citecolor    = blue,
     linkcolor    = blue,
     urlcolor     = blue
} % update hyperref color setting by LRZ

\begin{document}
\title[Berry curvature and edge states]{Berry Curvature and Bulk-Boundary Correspondence from Transport Measurement
for Photonic Chern Bands}

\author{Chao Chen}
\altaffiliation{These authors contributed equally to this work.}
\affiliation{Hefei National Laboratory for Physical Sciences at Microscale and Department of Modern Physics,
University of Science and Technology of China, Hefei, Anhui 230026, China}
\affiliation{CAS Centre for Excellence and Synergetic Innovation Centre in Quantum Information and Quantum Physics,
University of Science and Technology of China, Shanghai 201315, China}
\affiliation{National Laboratory of Solid State Microstructures, School of Physics, Nanjing University, Nanjing 210093, China}

\author{Run-Ze Liu}
\altaffiliation{These authors contributed equally to this work.}
\affiliation{Hefei National Laboratory for Physical Sciences at Microscale and Department of Modern Physics,
University of Science and Technology of China, Hefei, Anhui 230026, China}
\affiliation{CAS Centre for Excellence and Synergetic Innovation Centre in Quantum Information and Quantum Physics,
University of Science and Technology of China, Shanghai 201315, China}

\author{Jizhou Wu}
\altaffiliation{These authors contributed equally to this work.}
\affiliation{Department of Physics, Southern University of Science and Technology, Shenzhen, 518055, China}

\author{Zu-En Su}
\affiliation{The Physics Department and the Solid State Institute, Technion—Israel Institute of Technology, Haifa 3200003, Israel}

\author{Xing Ding}
\affiliation{Hefei National Laboratory for Physical Sciences at Microscale and Department of Modern Physics,
University of Science and Technology of China, Hefei, Anhui 230026, China}
\affiliation{CAS Centre for Excellence and Synergetic Innovation Centre in Quantum Information and Quantum Physics,
University of Science and Technology of China, Shanghai 201315, China}

\author{Jian Qin}
\affiliation{Hefei National Laboratory for Physical Sciences at Microscale and Department of Modern Physics,
University of Science and Technology of China, Hefei, Anhui 230026, China}
\affiliation{CAS Centre for Excellence and Synergetic Innovation Centre in Quantum Information and Quantum Physics,
University of Science and Technology of China, Shanghai 201315, China}

\author{Lin Wang}
\affiliation{Department of Physics, University of Konstanz, D-78457 Konstanz, Germany}

\author{Wei-Wei Zhang}
\affiliation{School of Computer Science, Northwestern Polytechnical University, Xi'an 710129, China}

\author{Yu He}
\affiliation{Shenzhen Institute for Quantum Science and Engineering, Southern University of Science and Technology, Shenzhen 518055, China}

\author{Xi-Lin Wang}
\affiliation{National Laboratory of Solid State Microstructures, School of Physics, Nanjing University, Nanjing 210093, China}

\author{Chao-Yang Lu}
\affiliation{Hefei National Laboratory for Physical Sciences at Microscale and Department of Modern Physics,
University of Science and Technology of China, Hefei, Anhui 230026, China}
\affiliation{CAS Centre for Excellence and Synergetic Innovation Centre in Quantum Information and Quantum Physics,
University of Science and Technology of China, Shanghai 201315, China}

\author{Li Li}
\email{eidos@ustc.edu.cn}
\affiliation{Hefei National Laboratory for Physical Sciences at Microscale and Department of Modern Physics,
University of Science and Technology of China, Hefei, Anhui 230026, China}
\affiliation{CAS Centre for Excellence and Synergetic Innovation Centre in Quantum Information and Quantum Physics,
University of Science and Technology of China, Shanghai 201315, China}

\author{Barry C. Sanders}
\email{sandersb@ucalgary.ca}
\affiliation{Hefei National Laboratory for Physical Sciences at Microscale and Department of Modern Physics,
University of Science and Technology of China, Hefei, Anhui 230026, China}
\affiliation{CAS Centre for Excellence and Synergetic Innovation Centre in Quantum Information and Quantum Physics,
University of Science and Technology of China, Shanghai 201315, China}
\affiliation{Institute for Quantum Science and Technology, University of Calgary, Alberta T2N 1N4, Canada}

\author{Xiong-Jun Liu}
\email{xiongjunliu@pku.edu.cn}
\affiliation{International Center for Quantum Materials, School of Physics, Peking University, Beijing 100871, China}
\affiliation{CAS Center for Excellence in Topological Quantum Computation, University of Chinese Academy of Sciences, Beijing 100190, China}
\affiliation{International Quantum Academy, Shenzhen 518048, China}

\author{Jian-Wei Pan}
\affiliation{Hefei National Laboratory for Physical Sciences at Microscale and Department of Modern Physics,
University of Science and Technology of China, Hefei, Anhui 230026, China}
\affiliation{CAS Centre for Excellence and Synergetic Innovation Centre in Quantum Information and Quantum Physics,
University of Science and Technology of China, Shanghai 201315, China}

%\date{\today}

\begin{abstract}
Berry curvature is a fundamental element to characterize topological quantum physics, while a full measurement of Berry curvature in momentum space was not reported for topological states. Here we achieve two-dimensional Berry curvature
reconstruction in a photonic quantum anomalous Hall system via Hall transport measurement of a momentum-resolved wave packet. Integrating measured Berry curvature over the two-dimensional Brillouin zone, we obtain Chern numbers corresponding to $-1$ and 0. Further, we identify bulk-boundary correspondence by measuring topology-linked chiral edge states at the boundary. The full topological characterization of photonic Chern bands from Berry curvature, Chern number, and edge transport measurements enables our photonic system to serve as a versatile platform for further
in-depth study of novel topological physics.
\end{abstract}

\maketitle

Being a central measure of the local geometry of Bloch bands, Berry curvature is a basic element to characterize topological phases~\cite{hasanColloquiumTopologicalInsulator2010,qiTopologicalInsulatorsSuperconductors2011}, and has a direct effect on wave-packet dynamics~\cite{xiaoBerryPhaseEffects2010a, nagaosaAnomalousHallEffect2010a}. Integrating Berry curvature over the Brillouin zone, one acquires the Chern number of the bulk, which is related to the gapless states at the boundary through the celebrated bulk-boundary correspondence. In solids, the boundary states may have novel transport effects~\cite{konigQuantumSpinHall2008,klitzingNewMethodHighAccuracy1980,zhangExperimentalObservationQuantum2005,changExperimentalObservationQuantum2013,dengQuantumAnomalousHall2020} and can be measured directly by angle-resolved photoemission spectroscopy~\cite{hsiehtopologicalDiracinsulator2008,liuDiscoveryThreeDimensionalTopological2014}.
For topological quantum simulation systems, including those with cold atoms, photonics, and solid-state qubits, various techniques have been developed for bulk-topology measurements, such as bulk spin textures~\cite{liuDetectingTopologicalPhases2013,wuRealizationTwodimensionalSpinorbit2016,songObservationsymmetryprotectedtopological2018,sunUncoverTopologyQuantum2018}, Berry phase~\cite{abaninInterferometricApproachMeasuring2013,atalaDirectmeasurementZak2013}, winding (or Chern numbers)~\cite{flurinObservingTopologicalInvariants2017,xuMeasuringWindingNumber2018,jotzuExperimentalRealizationTopological2014,aidelsburgerMeasuringChernNumber2015,mittalMeasurementTopologicalInvariants2016,tarnowskiMeasuringTopologyDynamics2019} and Berry curvature (or flux)~\cite{flaschnerExperimentalReconstructionBerry2016,wimmerExperimentalmeasurementBerry2017,ducaAharonovBohmInterferometerDetermining2015,gianfrateMeasurementquantumgeometric2020},
and edge-state measurements have also been reported~\cite{wangObservationunidirectionalbackscatteringimmune2009,rechtsmanPhotonicFloquetTopological2013,hafeziImagingtopologicaledge2013,manciniObservationchiraledge2015,stuhlVisualizingedgestates2015,xiaoNonHermitianbulkboundary2020}.
However,
all these studies suffer limitations:
in particular,
despite having been
applied to trivial systems,
Berry curvature reconstruction has not been reported for any topological system. Furthermore, although
edge modes and bulk topology were measured in a synthetic quantum Hall ribbon~\cite{chalopinProbingChiralEdge2020}, demonstrating bulk-boundary correspondence in more flexible lattices and broader topological systems is not yet reported but of great importance.

Here we first reconstruct Berry curvature over the two-dimensional (2D) Brillouin zone (BZ) for a photonic quantum anomalous Hall system via Hall transport measurement. We made state-of-the-art improvements to achieve this. First, we prepare a Gaussian wave packet with tunable central momentum by combining 2D time-bin encoding and phase-shifted nonunitary evolution.
Then we implement a two-loop-per-step 2D quantum walk (2DQW) evolution with rich Chern numbers on the momentum-scanned wave packet and observe its Hall transport under an effective external force to reconstruct Berry
curvature~\cite{xiaoBerryPhaseEffects2010a}. The 2D momentum-resolved measurement we achieved by observing wave-packet evolution was challenging in previous photonic experiments exploring topological edge-state dynamics with spatially localized initial states~\cite{rechtsmanPhotonicFloquetTopological2013,hafeziImagingtopologicaledge2013, chenObservationTopologicallyProtected2018,xiaoNonHermitianbulkboundary2020, chenTopologicalSpinTexture2022}, where all momentum components were mixed and unextractable.

Integrating the measured Berry curvature over the BZ, we obtain Chern numbers of~$0$ and $-1$.
Furthermore, we confirm bulk-boundary correspondence by observing chiral edge states at the interface, whose bulks on two sides are engineered to possess distinct Chern numbers. Therefore, our experiment provides a flexible photonic quantum simulation platform where both momentum-resolved transport measurement and edge-state measurement are feasible, which enables further in-depth study of novel topological physics.

\begin{figure}
\includegraphics[width=\columnwidth]{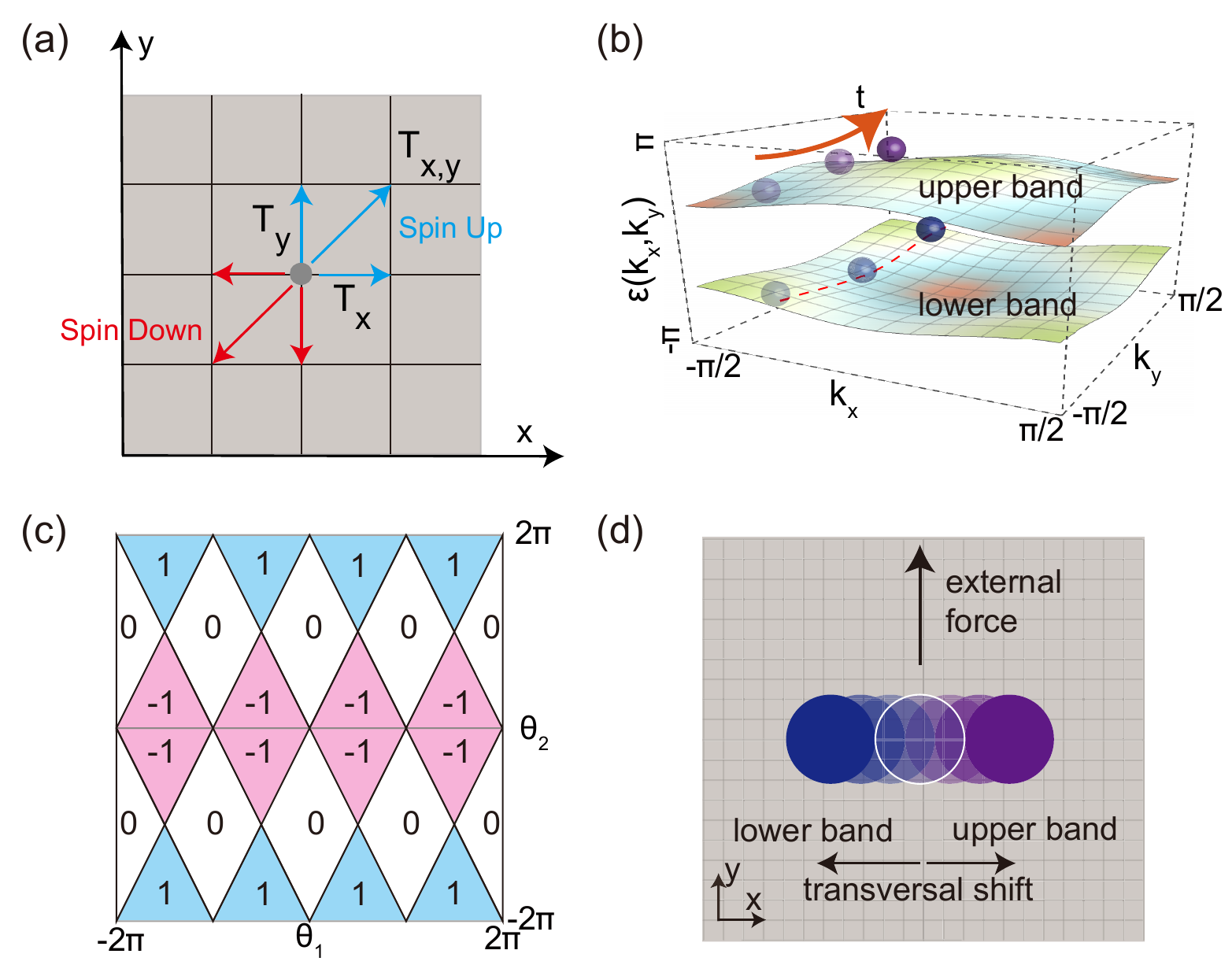}
\caption{Translation operations in 2DQW, energy band spectrum, phase diagram, and transversal Hall shift.
(a) Spin-dependent translations $T_x$, $T_y$ and $T_{xy}$. The walker of spin up and down shifts in the opposite directions.
(b)~Typical open-gap two-band quasienergy spectrum.
An external force along the~$y$ direction drives the eigenstate wave packets~(balls) moving in the $k_y$ direction.
(c)~Phase diagram of 2DQW in the parameter space $(\theta_1,\theta_2)$.
Different topological phases are denoted by the Chern number of~$\pm1$ and~$0$ for the lower band.
(d) Transversal wave-packet drift when the external force is along the~$y$ direction. The wave packets of the lower and upper bands drift in the opposite directions.}
\label{fig:principle}
\end{figure}

\begin{figure*}
\includegraphics[width=2\columnwidth]{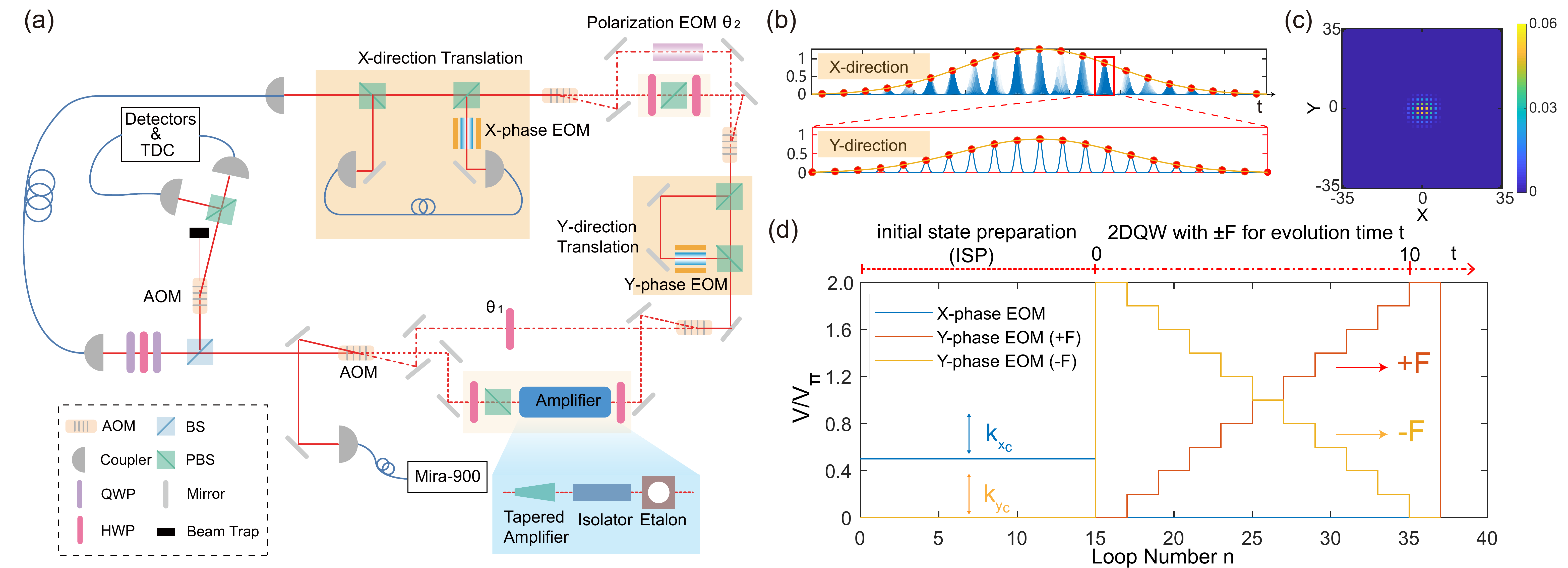}
\caption{Experimental setup, temporal-spatial coordinate mapping and electro-optical modulator signals.
(a)~Experimental setup of time-bin encoded 2DQW.
A picosecond laser pulse of \qty{905}{nm} generated by Mira-900 is injected into the circuit by an AOM.
Spin-dependent translation in the~$x$ ($y$) direction is implemented by a two-polarizing-beam-splitter fiber~(free-space) optical delay.
Between two translations, the light is switched by pairs of AOMs either for ISP~(dashed red line) or for 2DQW evolution~(dot-dashed red line).
(b)~Sketch of laser pulse distribution in temporal modes after the ISP.
(c)~Mapping the temporal distribution in (b)~to 2D position space. (d) Schematic phase EOM voltage signals $V$ versus loop number~$n$. The evolution time~$t$ is labeled as~$0$ when 2DQW starts after 15 loops' ISP and increases by~$1$ for every two loops.
During the 2DQW, an upward and downward ladder-type modulation is applied on the $y$-phase EOM
to implement an effective force in the~$\pm y$ direction, whose strength $F$ is tuned by the two-step difference.
The initial~$\bm{k}_{\text{c}}$ is tuned by voltages during ISP. $V_{\pi}$ is the half wave voltage.}
\label{fig:setup}
\end{figure*}

A 2DQW is programmable for quantum simulation~\cite{kitagawaExploringTopologicalPhases2010,schreiber2DQuantumWalk2012,flurinObservingTopologicalInvariants2017,barkhofenSupersymmetricPolarizationAnomaly2018,xuMeasuringWindingNumber2018,chalabiSyntheticGaugeField2019,wangSimulatingDynamicQuantum2019,geraldiExperimentalInvestigationSuperdiffusion2019,weidemannTopologicalFunnelingLight2020,quDeterministicSearchStar2022,chenTopologicalSpinTexture2022}.
Our model involves a spin-$\nicefrac{1}{2}$ particle in a periodic square lattice under the repeated unitary operation
\begin{equation}
U(\bm{\theta})
=T_x R(\theta_2)T_y R(\theta_1) T_{xy}R(\theta_1),\,
R(\theta)=\text{e}^{-\text{i}\frac{\theta}{2}\sigma_y},
\label{eq:1step}
\end{equation}
with the~$R$ spin rotation parametrized by $\bm{\theta}=(\theta_1,\theta_2)$
and~$\sigma_y$ a Pauli matrix.
Here
\begin{equation*}
%\label{eq:T_bullet}
T_\alpha=\sum_{\alpha\in \set{x,y,xy}}
\left[
\ket{\alpha+1}\!\bra{\alpha}
\otimes \ket{\uparrow}\!\bra{\uparrow}
+\ket{\alpha-1}\!\bra{\alpha}\otimes \ket\downarrow\!\bra\downarrow\right]
\end{equation*}
is a spin-dependent translation
in the $\alpha\in\{x,y,xy\}$ direction
as shown in Fig.~\ref{fig:principle}(a). To map out the Bloch band of $U(\bm{\theta})$,
we use Floquet band theory to get the effective Hamiltonian~$H(\bm{k})$ (see Supplemental Material \cite{supplemental_material} ).
The two-band quasienergy spectrum $\epsilon(\bm{k})$
over the 2D BZ
is shown in Fig.~\ref{fig:principle}(b). By varying $\bm{\theta}$,
the topological invariant---Chern number---with respect to the lower band can be controlled to be
$\pm1$ or~$0$, as the phase diagram Fig.~\ref{fig:principle}(c) shows.

We resolve the bulk topology by measuring transversal Hall drifts of a wave packet. We initially prepare a Gaussian wave packet
in position space, which evolves under unitary steps~Eq.~\eqref{eq:1step} with
applied external force $\bm{F}=F\hat{\bm{y}}$
(mimicking an electric force).
Given the force arising from a time-dependent vector potential, the crystal momentum is then modulated from~$\bm{k}$ to $\bm{k}+\nicefrac{\bm{F}\delta t}{\hbar}$ after time
$\delta t=t_\text{f}-t_0$ with $t_0$ ($t_{\text f}$) the initial~(final) time. The wave packet moves by~$\nicefrac{F\delta t}{\hbar}$
in the $k_y$ direction (Fig.~\ref{fig:principle}(b)). Meanwhile, as shown in Fig.~\ref{fig:principle}(d), $(x(t),y(t))$ the center of mass~(c.m.) of the wave packet experiences a transversal Hall drift in the~$x$ direction~\cite{xiaoBerryPhaseEffects2010a,changBerryCurvatureOrbital2008}, %which is labeled as
namely,
\begin{align}
    \Delta x=\int_{k_y(t_0)}^{k_y(t_{\text f})}
    \mathrm{d}k_y
    \left.\left(\frac1{F}\frac{\partial \epsilon_\text{l}(\bm{k})}{\partial k_x}+\Omega_\text{l}(\bm{k})\right)\right\vert_{k_x=k_x(t_0)}
    \label{eq:deltax},
\end{align}
with $(k_x(t),k_y(t))$ the momentum-space c.m.\ at time~$t$, and the subscript l denoting the lower band of interest.
The Berry curvature is
\begin{equation}
 \Omega_\text{l}(\bm{k})=\frac{\partial}{\partial k_x}\Braket{ u_\text{l}|\mathrm{i}\frac{\partial}{\partial k_y}|u_\text{l}}-\frac{\partial}{\partial k_y}\Braket{ u_\text{l}|\mathrm{i}\frac{\partial}{\partial k_x}|u_\text{l}}
\end{equation}
with $\Ket{u_\text{l}}$ the lower-energy eigenstate. Eq.~(\ref{eq:deltax}) shows that, except for energy-band dispersion, Berry curvature also gives rise to transversal drift,
which is known as the anomalous drift.

To extract Berry curvature from the Hall drifts, we need to cancel the contribution from the band dispersion.
Specifically, to measure Berry curvature around a momentum $\bm{k}_{\text{c}}:=(k_{x_{\text{c}}},k_{y_{\text{c}}})$,
we apply two
opposite forces $\pm F$ for time duration~$\delta t$. In these two cases,
wave packets initially centered at~$\bm{k}_{\text{c}}$ and $(k_{x_\text{c}},k_{y_\text{c}}+F\delta t)$,
respectively,
are driven over the same path in momentum space
but in opposite directions as
\begin{equation}
(k_{x_\text{c}},k_{y_\text{c}})\autoleftrightharpoons{$-F$}{$+F$} (k_{x_\text{c}},k_{y_\text{c}}+F\delta t). \end{equation}
Denoting the two displacements $\Delta x$ when~$\pm F$ are applied by~$\Delta x^\pm$,
half the difference is
\begin{equation}
   \Lambda= \frac{\Delta x^+-\Delta x^-}{2}
   =\int_{k_{y_{\text{c}}}}^{k_{y_{\text{c}}}+F\delta t}\text{d}k_y\Omega(k_{x_\text{c}},k_y)
    \label{eq:lambda},
\end{equation}
namely the integral only over Berry curvature (see Supplemental Material \cite{supplemental_material}). Therefore, Berry curvature can be extracted directly from transversal Hall drifts of the wave packet. For convenience, we always prepare wave packets with c.m.\ at a fixed site in position space,
denoted as~$(0,0)$,
so that we only need to measure $x^{\pm}(t_{\text f})$ to get~$\Lambda$.

Experimentally, we implement
2DQWs with photons in a laser pulse~\cite{schreiberPhotonsWalkingLine2010a,schreiber2DQuantumWalk2012,wimmerExperimentalmeasurementBerry2017}, and the coherence of the laser source suffices to achieve quantum interference
needed for photonic quantum walks \cite{peretsRealizationQuantumWalks2008};
of course,
two-particle or multiparticle quantum walks would need nonclassical sources providing, e.g., single photons as input~\cite{sansoniTwoParticleBosonicFermionicQuantum2012,crespiAndersonLocalizationEntangled2013,pouliosQuantumWalksCorrelated2014a}. The 2D lattice is encoded in photons' time-bin modes,
where the position space~$x, y$ is represented by the temporal modes with time intervals of \qty{1.12}{ns} and \qty{34.06}{ns}, respectively. The horizontal~(vertical) polarization of photons defines spin up~(down) of the walker. The walker firstly goes through nonunitary evolution for initial state preparation
(ISP)~\cite{regensburgerPhotonPropagationDiscrete2011,wimmerOpticaldiametricdrive2013}, during which the considerable optical loss is compensated for by a tapered optical amplifier (see Fig.~\ref{fig:setup}(a) and Supplemental Material \cite{supplemental_material}).
After~15 loops for ISP,
mapping back the temporal laser pulse distribution (Fig.~\ref{fig:setup}(b)) to 2D position space, the walker has a Gaussian distribution~(Fig.~\ref{fig:setup}(c)),
which corresponds to a Gaussian wave packet with momentum distribution width $\Delta k\approx 0.095\pi$ (see Supplemental Material \cite{supplemental_material}).

After ISP, we switch two pairs of acousto-optic modulators~(AOMs) so that the walker starts 2DQW as Eq.~\eqref{eq:1step},
where a half wave plate (HWP)
and a polarization electro-optical modulator~(EOM)
rotate the light polarization as
$R(\theta_{1,2})$ (Fig.~\ref{fig:setup}(a)).
$T_{xy}$ is realized by implementing~$T_x$ directly after~$T_y$;
i.e., $T_{xy}=T_x T_y$. The effective forces~$\pm F$ along the $y$ direction are implemented when upward and downward ladder-type modulations are applied to the $y$ phase EOM, respectively~(see Fig.~\ref{fig:setup}(d) and Supplemental Material \cite{supplemental_material}).
Owing to $\sim$50\% optical loss per loop during
the 2DQW process,
the pulses survive at a single-photon level and are detected by single-photon detectors after traveling for 20 loops,
corresponding to ten walking steps
(i.e., $\delta t=10$).

To deterministically map out the Berry curvature and Chern number of the lower band,
we require that,
when scanning~$\bm{k}_{\text{c}}$ in the BZ,
wave-packet
transversal drifts for the lower band be measured.
As the spin eigenstate varies at different momenta,
the horizontally polarized initial wave packet could overlap with both bands.
In this case, we observe that the initial wave packet splits into two oppositely drifted parts,
respectively corresponding to upper and lower bands.
In experiments,
we choose~$\bm{\theta}$ such that the band gap is open.
Owing to continuity of band dispersion and Berry curvature,
when we scan~$\bm{k}_{\text{c}}$ in the BZ,
the c.m.\ of the evolved wave packet of the lower~(upper)
band is also continuous in position space.
Then, when our initial wave packet
is prepared to be dominantly overlapping with the lower band
at~$\bm{k}_{\text{c}}$,
we can track the evolved wave packets related to the lower band by continuity as we slowly change~$\bm{k}_{\text{c}}$
to a neighboring momentum site (see Supplemental Material \cite{supplemental_material}).

\begin{figure}
\includegraphics[width=\columnwidth]{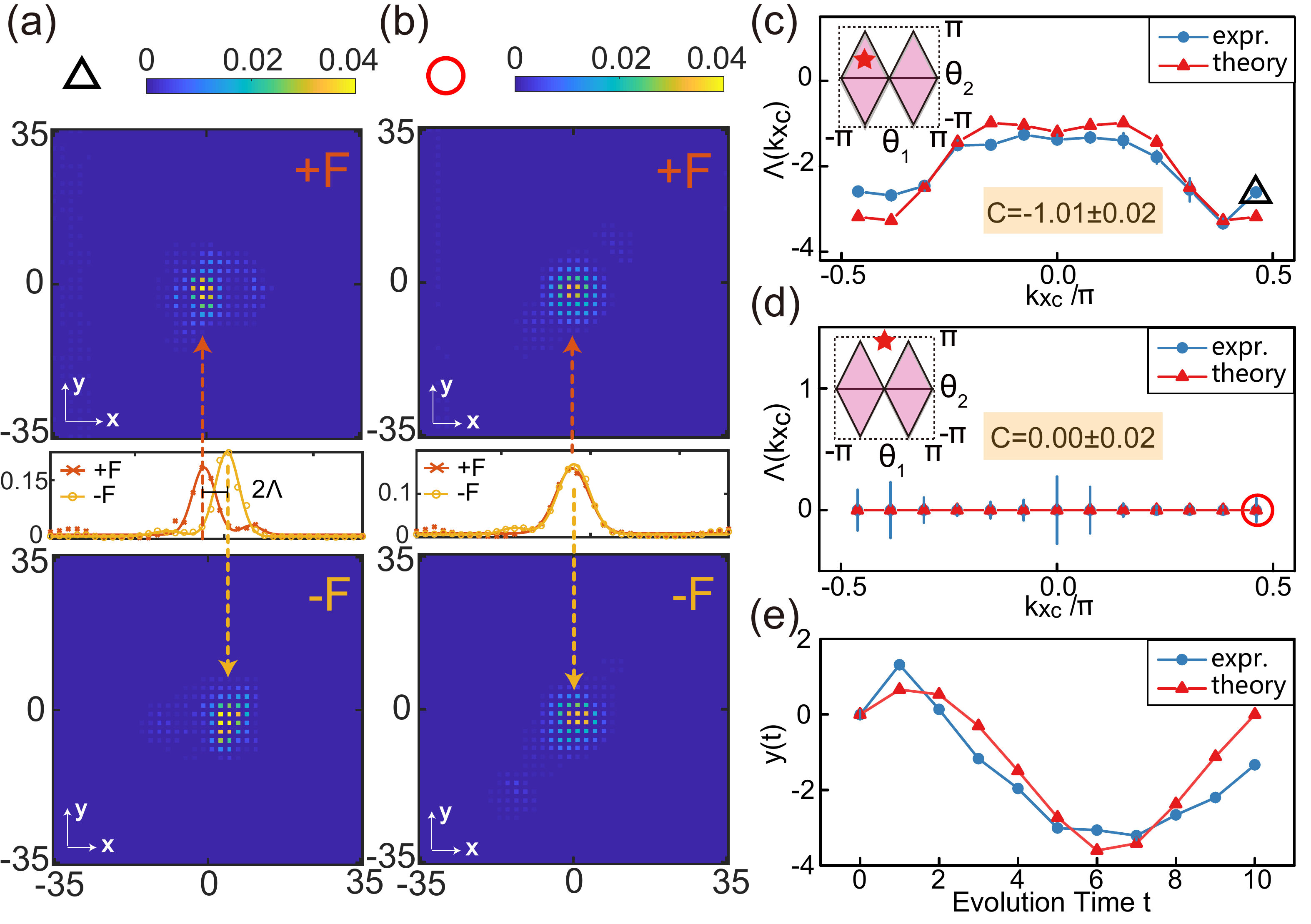}
\caption{Measured transversal Hall drifts, Chern number, and the wave-packet recurrence under the Bloch oscillation in $k_y$ with $F\delta t=\pi$. (a)~Measured probability distribution $P(x,y)$ after ten steps 2DQW under $\pm F$ forces. We set $\bm{\theta}=(-\nicefrac{\pi}{2},\nicefrac{\pi}{2})$ and $k_{x_\text{c}}\approx 0.46\pi$. The inset shows the marginal distributions $P(x)$ for the measured $P(x,y)$, where the distance between the wave-packet c.m.\ under~$\pm F$ is labeled.
(b)~Same as (a)~except for $\bm{\theta}=(0,\nicefrac{5\pi}{6})$. (c)~Measured (simulated) $\Lambda(k_{x_\text{c}})$ versus $k_{x_\text{c}}$ with $\bm{\theta}$ of (a). The $\Lambda(k_{x_\text{c}})$ marked by the triangle is obtained from the displacement shown in (a). By summing up $\Lambda(k_{x_\text{c}})$, we get the Chern number $-1.01\pm 0.02$. (d) Same as (c)~except for the $\bm{\theta}$ of (b). In this case, $C=0.00\pm 0.02$ is obtained. The $\Lambda(k_{x_\text{c}})$ marked by the red circle corresponds to the situation shown in (b).
(e) Recurrence of the wave packet in the~$y$ direction during the Bloch oscillation when $\bm{\theta}$ of (a) is chosen.
}
\label{fig:exp}
\end{figure}

We first measure the Chern number by discrete Bloch oscillations,
with~$k_y$ driven over the whole BZ under the external force ($F\delta t=\pi$) while $k_x$ is scanned discretely. Parameters
$\bm{\theta}$ of $(-\nicefrac{\pi}{2},\nicefrac{\pi}{2})$ and $(0,\nicefrac{5\pi}{6})$
are chosen, whose effective Hamiltonians
are topologically distinct. The applied force ($F=\nicefrac{\pi}{10}$) is much smaller than the band gap $\Delta E\approx\nicefrac{\pi}{2}$. Typical probability distributions $P(x,y)$ of finding the walker at $(x,y)$ on the lattice after ten steps of the 2DQW with~$\pm F$ applied are shown in Figs.~\ref{fig:exp}(a) and~\ref{fig:exp} (b). Transversal Hall drifts in the~$x$ direction are observed.
By Gaussian fit to the marginal distribution $P(x)$, we obtain the distance between the $x^\pm(t_\text{f}=10)$. After the complete Bloch oscillation in the $k_y$ direction,
we have $\Lambda(k_{x_\text{c}})\approx\int_{k_{y_\text{c}}}^{k_{y_\text{c}}+\pi}\mathrm{d}k_y\Omega(k_{x_\text{c}},k_y)$, where $k_{y_\text{c}}$ is fixed to be 0. By summing up $\Lambda(k_{x_\text{c}})$ over $k_{x_\text{c}}\in [-\nicefrac{\pi}{2},\nicefrac{\pi}{2}]$,
we obtain the Chern number according to $C=\int_{\text{BZ}}\Omega(\bm{k})\mathrm{d}\bm{k}/2\pi$. As expected, the measured Chern numbers are $-1.01\pm0.02$ and $0.00\pm0.02$ for the topological regime and trivial regime shown in Figs.~\ref{fig:exp}(c) and (d), respectively. In addition, we observe the wave-packet recurrence in the~$y$ direction, the evidence of Bloch oscillation (see Fig.~\ref{fig:exp}(e)~and Supplemental Material \cite{supplemental_material}).

\begin{figure}
\includegraphics[width=\columnwidth]{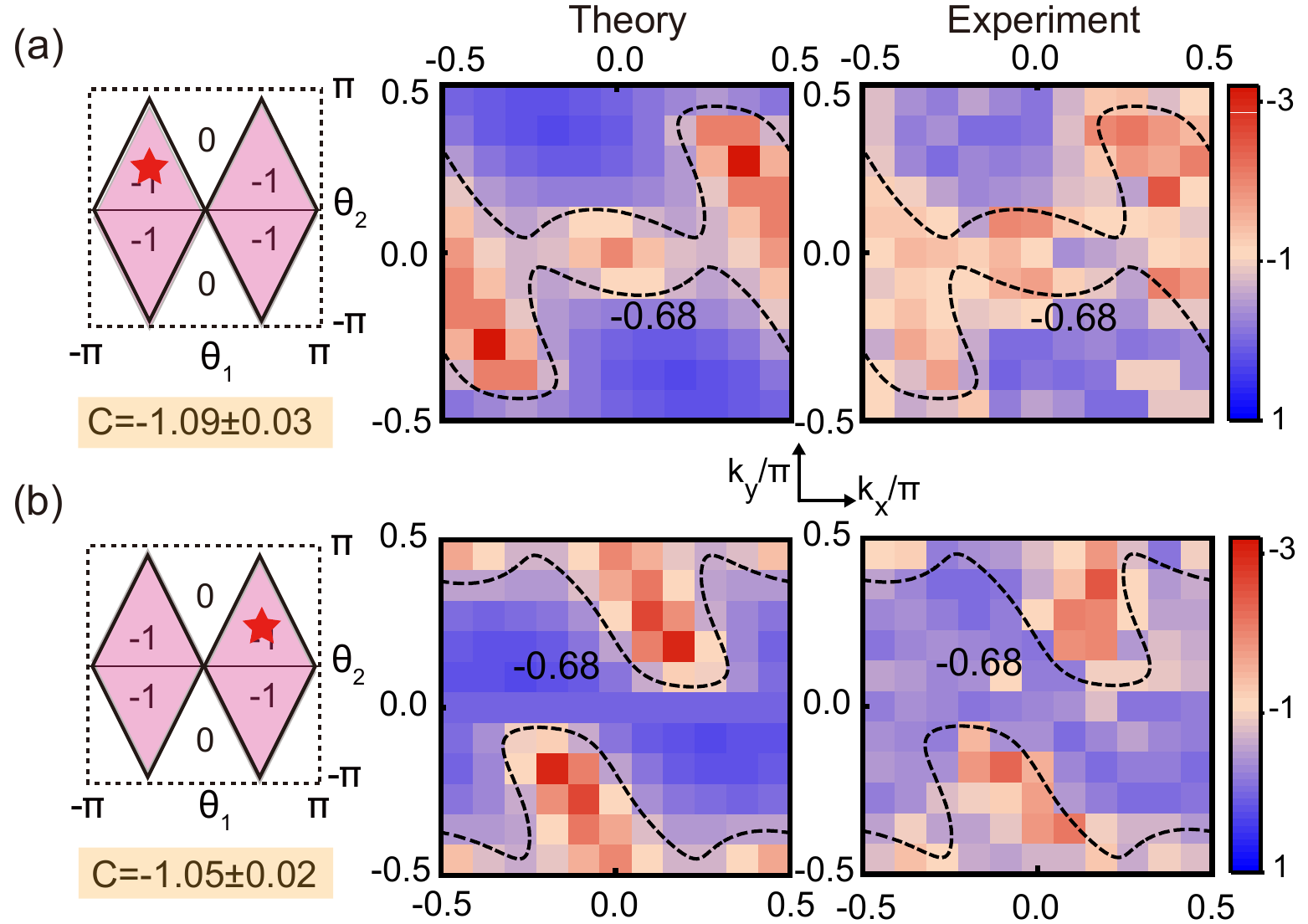}
\caption{Reconstruction of Berry curvature in the 2D Brillouin zone. (a) Theoretically simulated~(left) and measured~(right) Berry curvature when $\bm{\theta}=(-\nicefrac{\pi}{2},\nicefrac{\pi}{2})$ marked by the red star in the phase diagram. By scanning $\bm{k}_\text{c}$ in steps of $\nicefrac{\pi}{11}$ in the BZ,
we obtain the local Berry curvature $\Omega(\bm{k}_\text{c})$ from the measured $\Lambda(\bm{k}_\text{c})$. The contour lines of Berry curvature being -0.68 are a guide for the eye. Measured Berry curvature configurations agree with the theoretical ones in general. The momentum resolution can be promoted by using long-time initial-state evolution to prepare the Gaussian wave packet with smaller momentum distribution width. (b)~Theoretically simulated and measured Berry curvature when $\bm{\theta}=(\nicefrac{\pi}{2},\nicefrac{\pi}{2})$}.

\label{fig:berry cur}
\end{figure}

We then measure the complete landscape of Berry curvature in the 2D BZ and further confirm the bulk Chern numbers.
In this process, we set $F\delta t$ to be~$\nicefrac{\pi}5$,
which is driven by nine steps,
making the effective $F$ as small as $\nicefrac{\pi}{45}$. Small~$F$ facilitates the wave packet of one band to be separated from the other after the evolution, because the drift~$\Delta x$ is proportional to the inverse of $F$ (Eq.~\eqref{eq:deltax}). For a specific~$\bm{k}_{\text{c}}$, the measured~$\Lambda$ corresponds to the averaged local Berry curvature around~$\bm{k}_{\text{c}}$ over a distance of $\nicefrac{\pi}{5}$ in the $k_y$ direction.
By scanning~$\bm{k}_{\text{c}}$ over the whole BZ,
the entire Berry curvature landscape is reconstructed (see Supplemental Material \cite{supplemental_material}).
As shown in Fig.~\ref{fig:berry cur}, two different landscapes are measured corresponding to experimental parameters $\bm{\theta}=(\nicefrac{-\pi}{2},\nicefrac{\pi}{2})$ and $\bm{\theta}=(\nicefrac{\pi}{2},\nicefrac{\pi}{2})$.
When we sum up the measured Berry curvature shown in
Fig.~\ref{fig:berry cur}(a) and~\ref{fig:berry cur}(b) over the 2D BZ,
we obtain the integral Chern numbers as
$-1.09\pm 0.03$
and $-1.05\pm 0.02$.
This result matches well with the measurements by scanning
only $k_{x_\text{c}}$ (Fig.~\ref{fig:exp}),
showing that Berry curvature reconstruction is achieved for topologically nontrivial phases.

\begin{figure}
\includegraphics[width=\columnwidth]{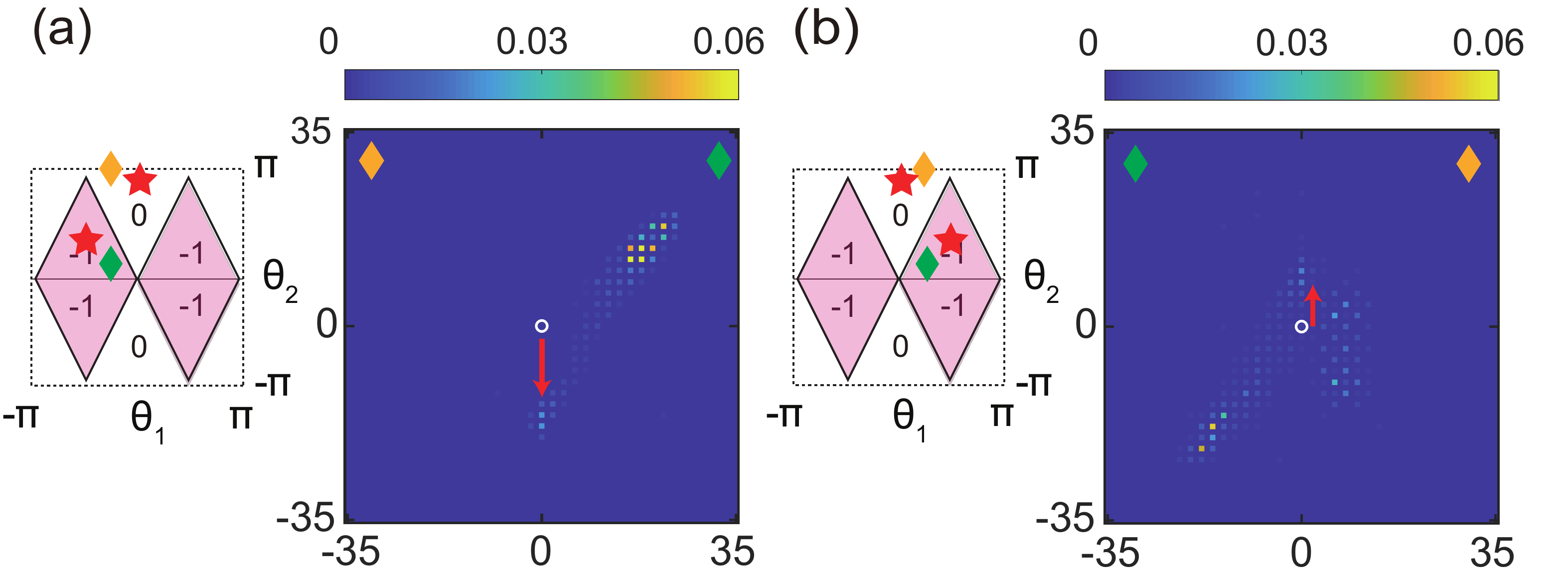}
\caption{Chiral edge states. (a) Measured probability distribution $P(x,y)$ after 12 steps' inhomogeneous 2DQW with an edge along the~$y$ direction between $x<0$ $(C=0)$ and $x\geq0$ $(C=-1)$.
Parameters $\bm{\theta}^{\text{left}}=(-\nicefrac{\pi}{4},\pi)$ (orange diamond) and $\bm{\theta}^{\text{right}}=(-\nicefrac{\pi}{4},\nicefrac{\pi}{5})$ (green diamond) are chosen. A spin-up polarized walker starts at site $(0,0)$ marked by a white circle. The red arrow denotes the propagation direction of the chiral edge states. (b)~Measured $P(x,y)$ when $\bm{\theta}^{\text{left}}=(\nicefrac{\pi}{4},\nicefrac{\pi}{5})$ and $\bm{\theta}^{\text{right}}=(\nicefrac{\pi}{4},\pi)$. Parameters for bulk topology measurement in Figs.~\ref{fig:exp} and \ref{fig:berry cur} are represented by red stars. }
\label{fig:edge states}
\end{figure}

Up to now, we have confirmed that the topologically nontrivial (trivial) bands of Chern number $C=-1~(0)$ can be constructed by tuning the parameters based on Berry curvature measurement. We further identify bulk-boundary correspondence that states the difference of the bulk Chern numbers on the two sides of an edge equals the number of edge modes existing on the edge. For this, we construct a boundary between the regions of $C=0$ and $C=-1$. The trivial and topological phase regimes are identical to those measured in Figs.~\ref{fig:exp} and~\ref{fig:berry cur};
however, because of the experimental limitation ($\theta_1$ must be the same for the left and right lattice) we choose the parameters of edge measurements to slightly deviate from those in the bulk measurements (see Fig.~\ref{fig:edge states} and the Supplemental Material \cite{supplemental_material}).

When initializing a walker on an edge where the bulk Chern number across varies (Fig.~\ref{fig:edge states}(a)), we find the walker to have a significant probability of propagating along the edge compared with the homogeneous 2DQW cases (see Supplemental Material \cite{supplemental_material}),  showing the presence of chiral edge states. The edge states propagate along the edge in the opposite direction when the topologies of the two sides are exchanged by tuning $\bm{\theta}$ (see Fig.~\ref{fig:edge states}(b) and Supplemental Material \cite{supplemental_material}). The reason that the walker also can be found either in the left or right bulks is because the initial states have some bulk-mode components~\cite{chenObservationTopologicallyProtected2018}. This observation is realized by directly switching the initial laser pulse into the 2DQW circuit without the Gaussian wave-packet preparation. Temporal-position mapping is reconstructed to create the boundary. The ability of measuring both the Berry curvature in the BZ and the chiral edge states in the position space directly
confirms the bulk-boundary correspondence, and is the key advantage of our experimental setup compared with other experimental quantum simulation systems.

In conclusion, we fully characterize photonic Chern bands by reconstructing Berry curvature through Hall drift, and measuring Chern number and chiral edge states, with which the bulk-boundary correspondence is experimentally substantiated. Our achievement of the momentum-resolved transport measurement together with edge-state measurement in a programmable 2D lattice makes our photonic time-bin system a versatile platform to investigate broad topological physics including anomalous Floquet topological states. The Berry curvature reconstruction with transport measurement may also enable a systematic study in precisely
revealing all the different micromechanisms of anomalous Hall transport by flexibly engineering disorder in the
present photonic lattice, including the Berry curvature
and disorder effects \cite{nagaosaAnomalousHallEffect2010a}, which cannot be directly observed
in solid systems. %in the present photonic lattice the onsite disorder can be flexibly engineered. This
Our system may be further extended to simulate other topological systems with various symmetries and dimensions~\cite{kitagawaExploringTopologicalPhases2010} and non-Hermitian topological systems with real gain and loss~\cite{longhiSelfHealingNonHermitianTopological2022}, under complicated experimental conditions, like electromagnetic fields~\cite{cedzichPropagationQuantumWalks2013,sajidCreatinganomalousFloquet2019}, and incommensurate potentials~\cite{wangRealizationDetectionNonergodic2020a}. %or disorder~\cite{nagaosaAnomalousHallEffect2010a}.
\\[0.1pt]
\begin{acknowledgements}
We thank J. Y. Zhang, D. W. Wang, Z. Y. Wang, X. C. Cheng, and M. C. Chen for helpful discussions, and Y. H. Li and J. Yin for providing a vacant optical table. Our work is supported by the Chinese Academy of Sciences (Grant No.\ XDB28000000), the Science and Technology Commission of Shanghai Municipality, the National Key R\&D Program of China (Grant No.\ 2021YFA1400900), National Natural Science Foundation of China (Grant Nos.\ 11825401, 11921005, 12261160368, 12104101), the Innovation Program for Quantum Science and Technology (Grant No. 2021ZD0302000), the Fundamental Research Funds for the Central Universities,  the Shenzhen Science and Technology Innovation Commission (Grant No.\ KQTD20200820113010023), China Postdoctoral Science Foundation (Grant No.\ 2023M731532) and Anhui Initiative in Quantum Information Technologies.
\end{acknowledgements}

~\\

\noindent
{\it Note added.}—Recently, we became aware of a related work
extracting Berry curvature by Bloch-state tomography~\cite{yiExtractingQuantumGeometric2023}.

%\bibliographystyle{apsrev4-1}
%\bibliography{berry_cur}

%

\end{document}